\documentclass[10pt]{article}
\usepackage{fancyhdr}
\usepackage{extramarks}
\usepackage{amsmath}
\usepackage{amsthm}
\usepackage{amsfonts}
\usepackage{siunitx}
\usepackage{tikz}
\usepackage[plain]{algorithm}
\usepackage{algpseudocode}
\usepackage{multirow}
\usepackage{booktabs}
\usepackage{palatino}
\usepackage{graphicx}
\usepackage{subfigure}
\usepackage[colorlinks,linkcolor=black,anchorcolor=black,citecolor=black,urlcolor=blue]{hyperref}
\usepackage{amsmath,bm}
\usepackage{booktabs}
\usepackage{mathtools}
\usepackage{amssymb}
\usepackage{caption}
\usepackage{capt-of}
\usepackage{mciteplus}
\usepackage{cite}
\usepackage{mathrsfs}
\usepackage[title,titletoc,toc]{appendix}
\usepackage{xr}
\usepackage{parskip}
\usepackage{soul}
\usepackage{textcomp}
\usepackage[colaction]{multicol}
\usepackage[switch]{lineno}
\usepackage{lipsum}
\usepackage{etoolbox}
\usepackage{longtable}
\usepackage{array}
\usepackage{tablefootnote}
\newcolumntype{C}[1]{>{\centering\arraybackslash}p{#1}}
\usetikzlibrary{automata,positioning}
\usetikzlibrary{automata,positioning}

\begin{document}

\title{Decoding asymptomatic COVID-19 infection and transmission}

\author{Rui Wang$^1$, Jiahui Chen$^1$, Yuta Hozumi $^1$, Changchuan Yin
$^2$, 
and Guo-Wei Wei$^{1,3,4}$\footnote{
Corresponding author.		E-mail: weig@msu.edu} \\
$^1$ Department of Mathematics, \\
Michigan State University, MI 48824, USA.\\
$^2$ Department of Mathematics, Statistics, and Computer Science, \\
University of Illinois at Chicago, Chicago, IL 60607, USA\\
$^3$ Department of Electrical and Computer Engineering,\\
Michigan State University, MI 48824, USA. \\
$^4$ Department of Biochemistry and Molecular Biology,\\
Michigan State University, MI 48824, USA. \\
}
\date{} 

\maketitle

\begin{abstract}
Coronavirus disease 2019 (COVID-19) is a continuously devastating public health and the world economy. One of the major challenges in controlling the COVID-19 outbreak is its asymptomatic infection and transmission, which are elusive and defenseless in most situations. The pathogenicity and virulence of asymptomatic COVID-19 remain mysterious. Based on the genotyping of 20656 Severe Acute Respiratory Syndrome Coronavirus 2 (SARS-CoV-2) genome isolates, we reveal that asymptomatic infection is linked to SARS-CoV-2 11083G$>$T mutation, i.e., leucine (L) to phenylalanine (F) substitution at the residue 37 (L37F) of nonstructure protein 6 (NSP6). By analyzing the distribution of 11083G$>$T in various countries, we unveil that 11083G$>$T may correlate with the hypotoxicity of SARS-CoV-2. Moreover, we show a global decaying tendency of the 11083G$>$T mutation ratio indicating that 11083G$>$T hinders SARS-CoV-2 transmission capacity. Sequence alignment found both NSP6 and residue 37 neighborhoods are relatively conservative over a few coronaviral species, indicating their importance in regulating host cell autophagy to undermine innate cellular defense against viral infection. Using machine learning and topological data analysis, we demonstrate that mutation L37F has made NSP6 energetically less stable. The rigidity and flexibility index and several network models suggest that mutation L37F may have compromised the NSP6 function, leading to a relatively weak SARS-CoV subtype. This assessment is a good agreement with our genotyping of SARS-CoV-2 evolution and transmission across various countries and regions over the past few months. 
\end{abstract}
Key words: COVID-19, SARS-CoV-2, NSP6, autophagy,  hypotoxicity, algebraic topology, network theory, machine learning  
\pagenumbering{roman}
\begin{verbatim}
\end{verbatim}

%
  \newpage


\setcounter{page}{1}
\renewcommand{\thepage}{{\arabic{page}}}

\section{Introduction}
The ongoing global pandemic of coronavirus disease 2019 (COVID-19) caused by Severe Acute Respiratory Syndrome Coronavirus 2 (SARS-CoV-2) has spread to more than 215 countries and territories with 8385440 positive cases and 450686 fatalities as of June 19, 2020 \cite{who_2020}. 
 Unlike SARS-CoV that mainly infects the lower respiratory tract,  SARS-CoV-2 is observed with a high level of shedding at the upper respiratory tract     \cite{wolfel2020virological}. A wide variety of  COVID-19 symptoms have been reported, including fever or chills, body or muscle aches, headache, stuffy or congested nose, dry cough, fatigue, sore throat, and loss of taste or smell, 2-14 days after exposure to the virus \cite{wolfel2020virological}.  Severe symptoms like high fever, severe cough, and shortness of breath indicate the onset of pneumonia. Less common gastrointestinal symptoms like diarrhea, nausea, and vomiting have been listed at the \href{https://www.cdc.gov/coronavirus/2019-ncov/symptoms-testing/symptoms.html}{Centers for Disease Control and Prevention}. Currently, symptom-based testing, contact tracing, isolation, and quarantine are the main strategies for controlling and combating COVID-19. However,   viable viruses have also been isolated from asymptomatic cases \cite{oran2020prevalence}.  Asymptomatic and presymptomatic cases can play an important role in transmitting coronavirus \cite{furukawa2020evidence}. Elusive asymptomatic transmission is regarded as the Achilles’ heel of current strategies to control COVID-19 \cite{gandhi2020asymptomatic}. 
Early studies with 235 cases of influenza virus infections indicate that the duration of influenza viral shedding was shorter and decayed faster and quantitative viral loads were lower in  asymptomatic than in symptomatic cases. \cite{ip2017viral} The epidemiological and virological characteristics of COVID-19 asymptomatic  pathogenicity  remains a mystery.

SARS-CoV-2 is a non-segmented positive-sense RNA virus that belongs to the $\beta$-coronavirus genus, coronaviridae family, and   Nidovirales order.  Many RNA viruses, such as the flu virus,  are prone to mutations due to the lack of proofreading in their genetic evolutions.  Viral mutations are driven by a  variety of factors, including replication mechanism, polymerase fidelity,  access to proofreading or post-replicative repair, sequence context, cellular environment, and host immune responses or gene editing \cite{sanjuan2016mechanisms}. Benefit from an error-correction mechanism common to the Nidovirales order, the replication of coronavirus is regulated by a bi-functional enzyme nonstructure protein 14 (NSP14) \cite{ferron2018structural}. Therefore, SARS-CoV-2 maintains relatively high accuracy in virus replication and transcription compared with the flu virus. Nonetheless, SARS-CoV-2 has had 10620 single mutations for the genome collected on January 5, 2020 \cite{wu2020new,wang2020decoding}. Although the impacts of mutations on SARS-CoV-2 transmission and pathogenicity  \cite{forster2020phylogenetic,becerra2020sars} and COVID-19 diagnosis, vaccine, and medicine have been studied \cite{wang2020decoding},  little is known about the connection between viral evolution and asymptomatic transmissions.

This work reports the first association of a SARS-CoV-2 single nucleotide polymorphism (SNP) variant and COVID-19 asymptomatic cases based on the genotyping of 20656 SARS-CoV-2 genome isolates. We reveal a significant correlation between asymptomatic infections and  SARS-CoV-2  single mutation 11083G$>$T-(L37F) on NSP6.   NSP6 is a common protein of $\alpha$ and $\beta$-coronaviruses that locate at the endoplasmic reticulum (ER) \cite{cottam2014coronavirus}. As a multiple-spanning transmembrane protein, coronavirus NSP6 participates in viral autophagic regulation.  Autophagy degrades alien components to provide an innate defense against viral infection and promote cell death and morbidity. \cite{ouyang2012programmed}. In response to extreme cases of starvation, autophagy generates autophagosomes to transfer long-lived proteins, unnecessary or dysfunctional components to lysosomes for orderly degradation. Studies show that NSP6 undermines the capability of autophagosomes to transport viral components to lysosomes for degradation by rendering smaller diameter autophagosomes in nutrient-rich media and thus enhancing viral replication  \cite{cottam2014coronavirus}. Additionally, NSP6 proteins induce a higher number of autophagosomes per cell compared with starvation to facilitate the assembly of replicase proteins \cite{benvenuto2020evolutionary,cottam2014coronavirus}.  Although further studies are required to understand the molecular mechanism of the NSP6 regulation of autophagy, it is clear that coronavirus NSP6 is extremely important to viral protein folding, viral assembly, and the replication cycle.  It is of paramount importance to understand how NSP6 mutation L37F leads to COVID-19 asymptomatic transmission and reduced virulence. 

We show that  NSP6 is a relatively conservative protein and the region around residues is quite conservative over several SARS-CoV-2 related genomes to maintain the crucial regulative function of NSP6.  Using artificial intelligence (AI), topological data analysis (TDA), and a variety of network models,  we further demonstrate that mutation L37F disrupts the folding stability of NSP6. We uncover the correlation between NSP6 mutation L37F and weakened SARS-CoV-2 virulence. We also analyze the global transmission and find the decay tendency of  NSP6 mutation L37F.  We prove that the evolutionary dynamics of L37F is age- and gender-independent.     

\section{Results}
We analyze 20656 SARS-CoV-2 complete genome sequences deposited in GISAID (\url{https://www.gisaid.org/}) database up to June 19, 2020. Among them, 7243 samples have patient status information recorded as asymptomatic, symptomatic, hospitalized, ICU, deceased, and so on. In particular, 83 samples are labeled with asymptomatic (53)  and symptomatic (30) cases. By genotyping  53 genome samples, we find that mutation 1083G$>$T  is significantly relevant with asymptomatic infection with a p-value being much smaller than 0.05. 
Mutation 11083G$>$T changes leucine (L) residue at position 37 of NSP6 to phenylalanine (F), denoted as L37F. 
 
In the following sections, we  investigate   the relationship between mutation 11083G$>$T-(L37F)
and asymptomatic infections using both gene-specific analysis and protein-specific analysis. The global evolution and transmission pathway of the mutation is studied as well.

\subsection{Gene-specific analysis}

Our customized dataset generated from genotyping 20656 SARS-CoV-2 complete genome sequences downloaded from the GISAID database is summarized in Table \ref{tab:dataset}. Here, patient status referred to as incomplete records of asymptomatic, symptomatic, hospitalized, deceased, gender, etc. Additionally, our dataset contains data collection dates and locations of genome isolates.  The detailed information of our dataset is available in the Supporting information. 

\begin{table}[ht]
    \caption{Characteristics of customized dataset}
    \label{tab:dataset}
    \centering
    \begin{tabular}{ccccc}
    \hline
    Sample size & With 11803G$>$T & With patient status & Asymptomatic & Symptomatic\\ 
    \hline
    20656         & 2305  & 7243   & 53  & 30\\
    \hline
    \end{tabular}
\end{table}

We first analyze 83 sequences that have either asymptomatic or symptomatic labels submitted from Japan. The statistic analysis shows that the Pearson correlation coefficient between asymptomatic records and 10083G$>$T is 0.606. Moreover, the p-value is 1.29$\times10^{-9}$, which reveals the significant relevance between asymptomatic and single mutation 11083G$>$T-(L37F). Next, we split our genotyping dataset of 20656 sequences into different countries and extract those records with the 11083G$>$T mutation. Table \ref{tab:country ratio} summarizes the total number of sequences related to 11083G$>$T-(L37F), denoted as N$_{\text{L37F}}$, the total number of sequences N$_{\rm S}$, the 11083G$>$T-(L37F)  ratio, the number of total cases, the number of total deaths, and the death ratio of 25 countries up to June 19, 2020, respectively. A complete table for all countries/regions can be found in the Supporting information.
\begin{table}[ht]
    \centering
    \setlength\tabcolsep{5pt}
	\captionsetup{margin=0.1cm}
	\caption{The 11083G$>$T-(L37F) mutation ratio and death ratio in each country as of June 19,2020. Here, N$_{\text{L37F}}$ and N$_{\rm S}$ represent the total number of sequences with 11083G$>$T-(L37F) and the total number of sequences in each country listed in our dataset, respectively.}
    \label{tab:country ratio}
    \begin{tabular}{lcccccc}
    \hline
    Country/Region&N$_{\text{L37F}}$&N$_{\rm S}$&Mutation ratio &Total cases & Total deaths& Death ratio \\
    \hline
    Singapore&203&315&0.644&41473&26&0.06\%\\
    Japan&60&103&0.583&17740&935&5.27\%\\
    Turkey&24&66&0.364&184031&4882&2.66\%\\
    Jordan&7&22&0.318&1001&9&0.90\%\\
    India&166&540&0.307&380532&12573&3.30\%\\
    Norway&7&28&0.250&8629&244&2.83\%\\
    Australia&235&1141&0.206&7391&102&2.28\%\\
    South Korea&5&25&0.200&12306&280&2.28\%\\
    Iceland&77&425&0.181&1816&10&0.55\%\\
    United Kingdom&902&6037&0.149&300473&42288&14.07\%\\
    Thailand&7&58&0.121&3146&58&1.84\%\\
    Indonesia&1&9&0.111&42762&2339&5.47\%\\
    Canada&18&280&0.064&99853&8254&8.27\%\\
    Vietnam&3&47&0.064&342&0&0.00\%\\
    Belgium&33&599&0.055&60384&9683&16.04\%\\
    Italy&6&109&0.055&238159&34514&14.49\%\\
    South Africa&3&55&0.055&83890&1737&2.07\%\\
    China&15&284&0.053&84940&4645&5.47\%\\
    France&11&230&0.048&153557&29537&19.24\%\\
    United States&236&5835&0.04&2149166&117472&5.47\%\\
    Brazil&5&124&0.04&955377&46510&4.87\%\\
    Switzerland&7&180&0.039&31117&1677&5.39\%\\
    Spain&12&384&0.031&245268&27136&11.06\%\\
    Chile&5&161&0.031&225103&3814&1.69\%\\
    Russia&2&166&0.012&569063&7814&1.37\%\\
    \hline
    \end{tabular}
\end{table}

Table \ref{tab:country ratio} suggests that the 11083G$>$T-(L37F) ratio correlates with the death ratio. For example,  Singapore has the highest 11083G$>$T-(L37F) mutation ratio as of 0.644. A piece of recent news reported that half of the COVID-19 cases in Singapore are symptomless \cite{John_news}, which matches our finding that mutation 11083G$>$T-(L37F) is relevant to the asymptomatic infections. Moreover, Singapore has the second-lowest death ratio as listed in Table \ref{tab:country ratio}, suggesting that single mutation 11083G$>$T-(L37F) may have weakened SARS-CoV-2 virulence. A similar deduction can be obtained from the records of Turkey, Jordan, Norway, Australia, and South Korea. The relatively high 11083G$>$T-(L37F) mutation ratio (greater than 0.200) with a correspondingly low death ratio (less than 3.00\%) further validates that 11083G$>$T-(L37F) may be relevant to the asymptomatic-induced the hypotoxicity of SARS-CoV-2. Moreover, 11 out of 17 countries whose mutation ratios are less than 0.200 have a death ratio greater than 4.5\%. The death ratios of the United Kingdom, Belgium, France, and Spain are even higher than 10\%, which supports our assumption that mutation 11083G$>$T-(L37F) weakens SARS-CoV-2 virulence.  

\begin{figure}[ht]
    \centering
    \includegraphics[width=1.03\textwidth]{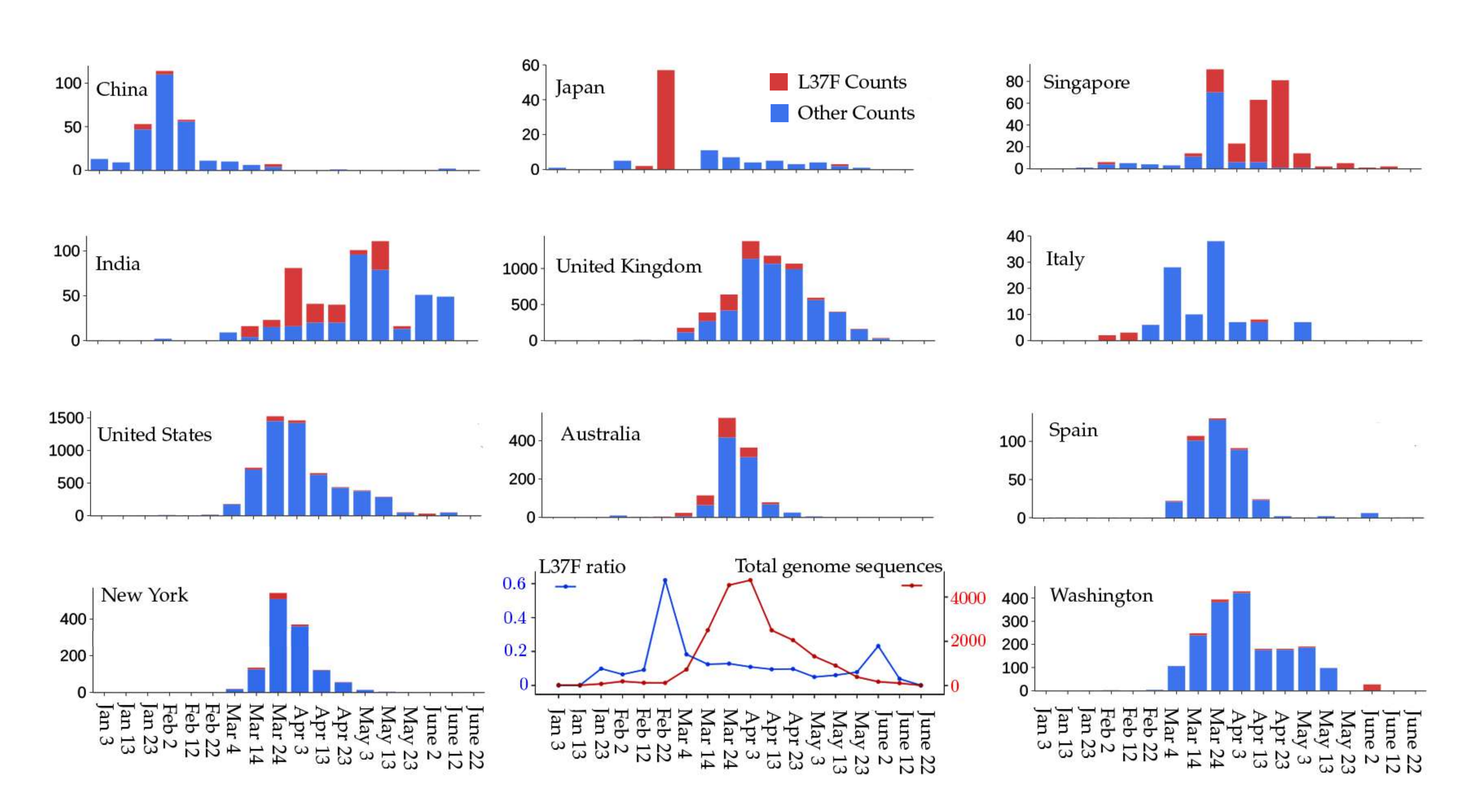}
    \caption{ The time evolution of SARS-CoV-2  mutation 11083G$>$T-(L37F) in nine countries and two states (i.e., New York and Washington) from the United States. The bar plots show the frequency of cases with and without 11083G$>$T-(L37F) mutation. Each bar width covers a 10-day period.   
The blue line plot illustrates the evolution of the L37F mutation ratio computed as the count of genome sequences having the L37F mutation over the total count of genome sequences at each time period.  The red line plot shows the evolution of the total count of genome sequences.
}
    \label{fig:Barplot}
\end{figure}

Figure \ref{fig:Barplot} illustrates the number of complete SARS-CoV-2 sequences in our dataset with 11083G$>$T-(L37F) detected versus the number of complete SARS-CoV-2 sequences without 11083G$>$T-(L37F) detected every 10 days in Singapore, Japan, India, China, United Kingdom, Italy, United States, Australia, and Spain, as well as two states in the United States: New York and Washington. The blue and red bar represents the 11083G$>$T-(L37F) counts and other mutation counts every ten-day period, respectively. Similar bar plots for all countries/regions  involving this specific mutation can be found in the Supporting information. Singapore's plot shows that 11083G$>$T was widely found after March 24, 2020, which is consistent with the report saying that at least half of Singapore's newly discovered COVID-19 cases show no symptoms \cite{John_news}. Almost all of the cases collected from February 12 to February 22 have 11083G$>$T mutation with asymptomatic infections recorded, which is the most robust evidence to associate the SARS-CoV-2 11083G$>$T or L37F mutation on NSP6 with the asymptomatic infections. India is one of the countries that also have a large number of  11083G$>$T mutation cases. Recent news reported that 80 percent of all coronavirus patients in India were asymptomatic or showed mild symptoms \cite{Desk_news}, which supports our deduction that 11083G$>$T correlates with the asymptomatic infection. Moreover, one can see that mutation 11083G$>$T in China, the United Kingdom, and Australia is not as abundant as in Singapore, Japan, or India. Furthermore, as discussed before, the asymptomatic infection may associate with the hypotoxicity of SARS-CoV-2 as well. Note that from Figure \ref{fig:Barplot}, China, the United States, Spain, and Italy have a relatively lower occurrence of 11083G$>$T, which may contribute to the relatively high death ratio.

However, this specific mutation is not the only factor in determining the death ratio. The number of total infected cases, diagnostic testing,  medical and health conditions,  age structure, nursing-home population, etc. are also critical factors. For example, Japan has a second-highest ratio of 11083G$>$T-(L37F), while the death ratio is greater than 5 percent.  Similarly, the United Kingdom has the third-highest death ratio, suggesting its ratio of mutation 11083G$>$T-(L37F) is not at the last echelon. One of the possible reasons is that the healthcare system was heavily broken by a sudden increase in infected cases. The lack of proper healthcare attention to a large number of mid/severe COVID-19 cases resulted in a high mortality ratio. Note that although Russia has the lowest 11083G$>$T-(L37F) mutation ratio lists in Table  \ref{tab:country ratio}, its 1.37\% death ratio caused by COVID-19 is also relatively low. One possible reason is Russia's relatively low median age.

In total, 53 genome samples were related to 11083G$>$T among 542 samples that were collected from May 14 to May 24 in New York. After that, fewer cases were collected in New York and the 11083G$>$T mutation ratio decreased. Also, the state of Washington had a small proportion of 11083G$>$T mutation before May 13, 2020. In the first week of June, 26 out of 27 genome sequences had mutation 11083G$>$T, which may indicate that the asymptomatic infection is becoming increasingly prevalent in Washington. 

During the last few months, the proportion of the 11083G$>$T mutation in the United Kingdom has gone down as shown in Figure \ref{fig:Barplot}.  Globally, the ratio of samples with the 11083G$>$T mutation over all samples, shown in the middle chart of the last row in  Figure \ref{fig:Barplot} decays over time, suggesting the mutation is unfavorable to the viral transmission.  Two abnormal peaks appeared on February 22 and June 2 in the ratio were due to the L37F mutation counts from Japan and Washington, respectively. The relatively small numbers of total genome sequences at these dates induce the jumps. Unfortunately, the number of genome sequences decays rapidly after April as shown in Figure \ref{fig:Barplot}, while globally, the number of SARS-CoV-2 infections increasing steadily according to the WHO's daily situation reports (\url{https://www.who.int/emergencies/diseases/novel-coronavirus-2019/situation-reports}). 

\subsection{Evolution and transmission analysis}

In this section, we track the global transmission pathways of the single mutation 11083G$>$T-(L37F) to understand its spread dynamics. We found that the first genome containing single mutation 11083G$>$T-(L37F) was sequenced in Chongqing, China, on January 18, 2020, which can be considered as the ancestor of 11083G$>$T in our dataset. However, the complete genome sequences released on the GISAID do not include all of the infected cases. Mutation 11083G$>$T  also detected in a co-mutation record [8782C$>$T, 11083G$>$T, 28144T$>$C] in Yunnan, China one day earlier than Chongqing's sequence, which indicates that the true ancestor of 11083G$>$T might have occurred early than January 18, 2020. Note that the first SARS-CoV-2 sequence released in GISAID was collected in Wuhan, China on December 24, 2019.  China's COVID19 epicenter, Wuhan, implemented a lockdown on January 23, 2020,  which explains why China has a relatively low ratio of 11083G$>$T-(L37F) mutation. 

The first confirmed case reported by the Singapore government was on January 23, 2020. Interestingly, the first genome sequence related to mutation 11083G$>$T-(L37F) was found in Singapore on January 29,2020, only a few days after the first confirmed cases. Therefore, the transmission of COVID-19 in Singapore at the initial stage contains a large proportion of the11083G$>$T-(L37F) mutation. The prevalence of the11083G$>$T-(L37F) mutation in Singapore's SARS-CoV-2 infections explains its low death ratio.

Mutation 11083G$>$T-(L37F) first appeared in the United States on January 22, 2020, in Arizona, a 26-year-old male with co-mutations [8782C$>$T, 11083G$>$T, 28144T$>$C, 29095C$>$T], which is the descendant of [8782C$>$T, 11083G$>$T, 28144T$>$C] found in China. Since then, more records related to 11083G$>$T have appeared in the United States. At this stage, 236 genome isolates from the United States in our dataset are relevant to 11083G$>$T. France and Italy are the following countries that found the 11083G$>$T mutation at an early stage. Both of them detected this mutation on Jan 29, 2020. Although 11083G$>$T mutation has arrived in the United States and the United Kingdom at a very early stage, different subtypes of SARS-CoV-2 spread vastly due to their large population, which makes the 11083G$>$T  mutation non-dominated. 

In Jordan, the earliest sequence released on GISAID is on March 16, 2020, with 11083G$>$T mutation, which indicates the high 11083G$>$T ratio in Jordan. Note that multiple reasons may lead to a low death ratio of 0.90\% in Jordan. One may be due to the hypotoxicity of SARS-CoV-2 caused by mutation 11083G$>$T. The other is that the healthcare system is in good condition due to the small number of infected patients.

The first confirmed COVID-19 cases reported by the Vietnamese government was on January 16, 2020. However, the first genome isolates related to mutation  11083G$>$T in our dataset was collected on March 17, 2020. The 11083G$>$T was found in Vietnam two months after the first confirmed cases, which result in a low 11083G$>$T mutation ratio. Surprisingly, none of the patients died in Vietnam. As a low-middle income country with nearly 100 million population and a much less-advanced healthcare system than other countries such as Italy, the United Kingdom, and France, Vietnam's success in tackling COVID-19 attracts attention. They immediately took action to ban the foreign nationals' entries and implement strict social distancing rules, which is the essential factor in keeping the coronavirus death toll at zero.

\begin{figure}[H]
    \centering
    \includegraphics[width=1\textwidth]{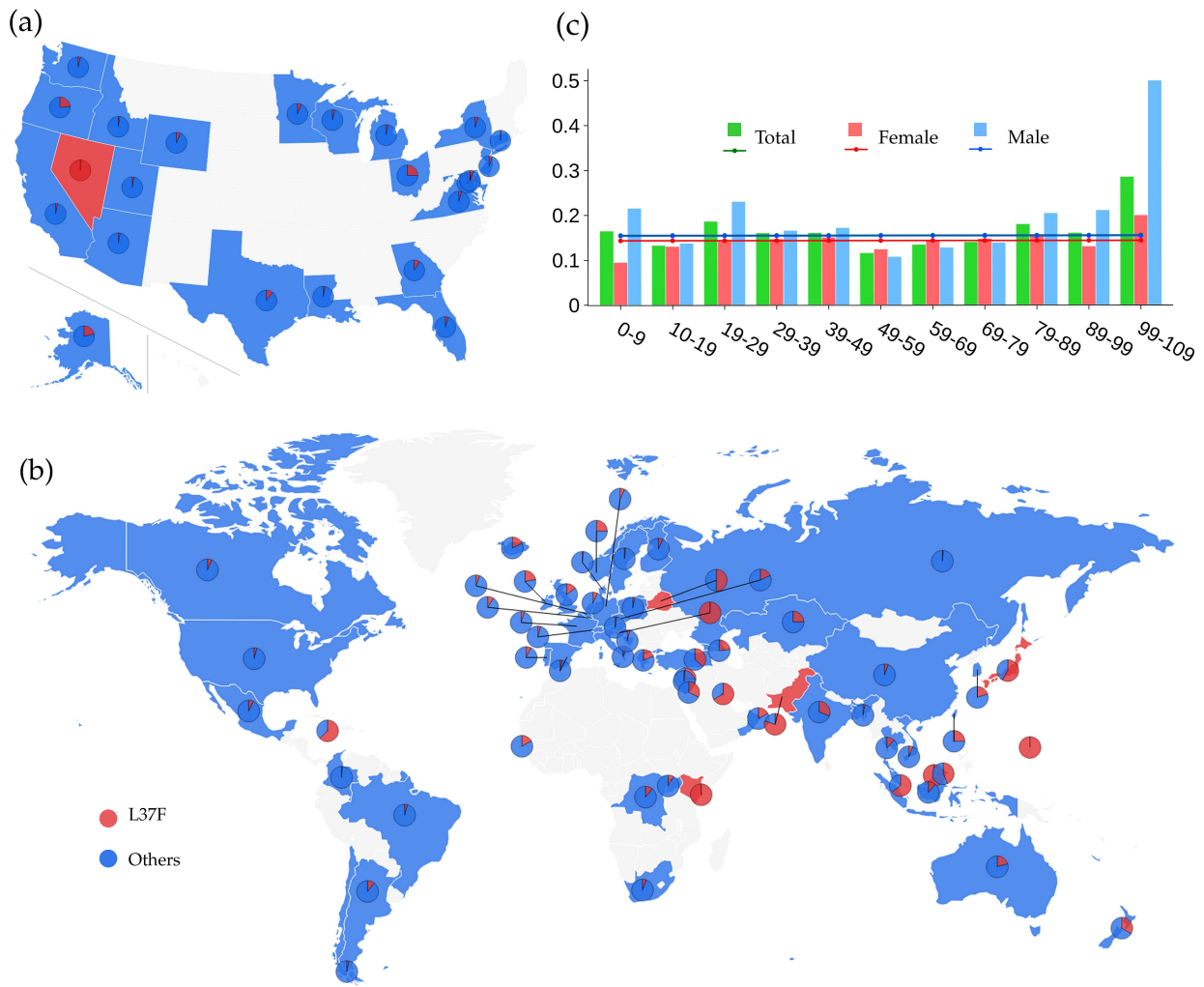}
    \caption{(a) 
Pie chart plot of the distribution of 
        genome samples with and without the 11083G$>$T-(L37F) mutation in the United States. The red and blue colors represent the 11083G$>$T-(L37F) mutation and non-11083G$>$T mutations, respectively. The color of the dominated mutation type decides the base color of each state. (b) Bar plot of the age and gender distributions of the ratio of the number of samples having mutation L37F over the total number of samples having age and/or gender labels. The straight lines over all age groups are the group average ratios (the average ratio from the total dataset having age labels overlaps with the average ratio from the dataset having male labels.).    
(c) Pie chart plot of the distribution of genome samples with and without the 11083G$>$T-(L37F) mutation in the world.  The color of the dominated mutation decides the base color of each country.}
    \label{fig:World}
\end{figure}

Figure \ref{fig:World} (a) generated by \href{https://www.highcharts.com/}{Highcharts} illustrates the distribution and proportion of 11083G$>$T-(L37F) mutation in the United States. The red and blue represent the 11083G$>$T-(L37F) mutation and non-11083G$>$T mutations. The color of the dominated type of mutation decides the base color of each state. One can see that Oregon, Alaska, Texas, and Ohio have a higher proportion of 11083G$>$T than the other states, while the death ratio in Oregon, Alaska, and Texas is lower than the average death ratio (5.47\%) in the United States as of June 19, 2020. Note that although the ratio of mutation 11083G$>$T in Nevada is 1, only one genome sequence was submitted from Nevada; hence, it is insignificant in the statistics.

 Figure \ref{fig:World} (b) created by using \href{https://www.highcharts.com/}{Highcharts} illustrates the distribution and proportion of  the 11083G$>$T-(L37F) mutation in the world. Mutation 11083G$>$T-(L37F) only dominated in a few countries such as Japan, Singapore, Pakistan, Jamaica, Kenya, Belarus, and Brunei. Most of the other countries have a small proportion of mutation 11083G$>$T-(L37F). Detailed information for all countries/regions is available in the Supporting information. 

Figure \ref{fig:World} (c) displays the potential age and gender disparities in mutation 11083G$>$T-(L37F). Among 20656 complete genome sequences,   9923 cases have age labels and 9437 cases (male cases: 4993 and female cases: 4444) have gender labels, which are used for our age-related and gender-related analysis, respectively. Overall, male cases have a slightly higher ratio of the 11083G$>$T-(L37F) mutation than female cases. In terms of age distribution, the ratio is significantly higher in the age group of 99-109 years. However, only 7 cases fall into this group and all data were collected from the United Kingdom (2 cases)  and Singapore (5 cases). More data and further studies are required to draw any conclusion from this group. Therefore, we conclude that mutation 11083G$>$T-(L37F) is quite evenly distributed over age and gender groups, implying that the mutation does not have a host-dependent behavior.  

\subsection{Protein-specific analysis} 

\begin{figure}[ht]
    \centering
    \includegraphics[width=0.95\textwidth]{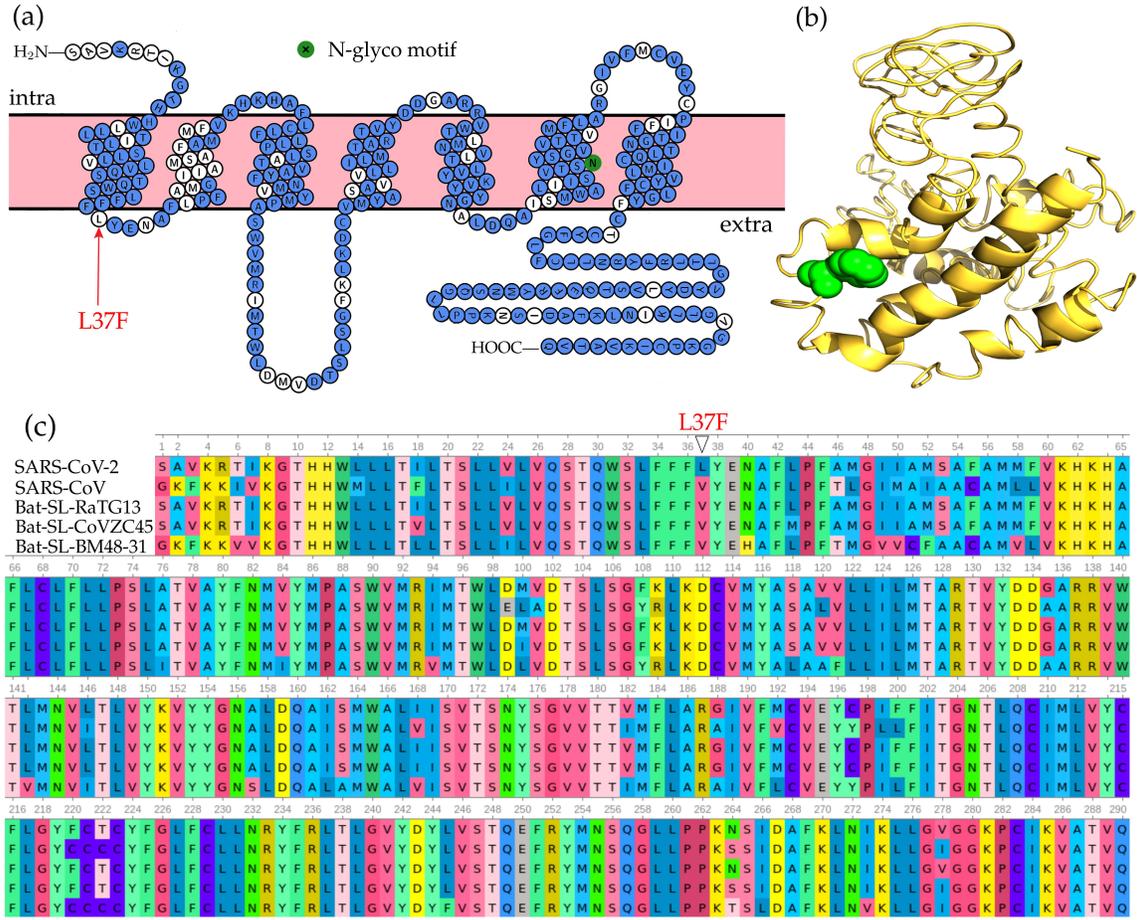}
    \caption{ (a) The visualization of SARS-CoV-2 NSP6 proteoform. The 11083G$>$T mutation in the genome sequence leads to the residue 37 leucine (L) mutant to phenylalanine (F). We use a red arrow to point out the mutation detected at residue 37. According to the sequence alignment results in (c), we color the conservative SARS-CoV-2 NSP6 residues by blue. (b) The 3D structure of SARS-CoV-2 NSP6 protein. Green represents the mutation residue at position 37 of NSP6. (c) Sequence alignments for the NSP6 proteins of  SARS-CoV-2, SARS-CoV, bat coronavirus RaTG13, bat coronavirus CoVZC45, bat coronavirus BM48-31.  The numbering is labeled according to SARS-CoV-2.
	}
    \label{fig:Protter}
\end{figure}

Figure \ref{fig:Protter} depicts the sequence and structural information of NSP6. Figure \ref{fig:Protter} (a) is generated by an online server \cite{omasits2014protter}, \autoref{fig:Protter} (b) is plotted by PyMol \cite{delano2002pymol}, and \autoref{fig:Protter} (c) is created by \cite{bah2007inkscape}. As shown in  Figure \ref{fig:Protter}(a), NSP6 is a multi-span membrane protein. In mutation 11083G$>$T-(L37F), both leucine and phenylalanine are $\alpha$-amino acids and non-polar. However, phenylalanine has a  benzyl ring on the side chain, making the secondary structure of NSP6 more compact. As shown in Figures \ref{fig:Protter}(a) and (c),  from the NSP6 protein residue position 34 to 37, the residues are FFFL. Therefore, there will be four continuous phenylalanine amino acid FFFF after the L37F mutation. As a result,  mutation L37F can stiffen the NSP6 structure by the aromatic-aromatic, hydrophobic, or $\pi$-stacking interactions. Such an additional interaction can significantly change NSP6 function \cite{benvenuto2020evolutionary}. 

Figure \ref{fig:Protter}(c) presents the alignment of the SARS-CoV-2 NSP6 sequence with those of the other four coronaviruses, including that of SARS-CoV.  The sequence identities between SARS-CoV-2 NSP6 and the NSP6s of SARS-CoV \cite{lee2003major}, RaTG13 \cite{zhou2020pneumonia},  ZC45\cite{hu2018genomic},  BM48-31 \cite{drexler2010genomic} are 87.2\%, 99.7\%, 97.9\%, and  83.8\%, respectively. Note that the overall sequence identity between SARS-CoV-2 and SARS-CoV is  83.8\%. Therefore, NSP6 is relatively conservative.  Additionally, NSP6 position 37 is in a relatively conservative environment among different species. Position 37 for all other species is valine, which is also hydrophobic and extremely similar to leucine. Two amino acids differ by one -CH$_2$- unit. A large region from 25 to 45 has only three mutations among five species, indicating the region is potentially important for the protein function.

There three-dimensional (3D) structure of NSP6 is displayed in Figure \ref{fig:Protter}(b). To further understand the impact of mutation L37F on NSP6, we carry several theoretical analyses using AI\cite{cang2017analysis}, TDA \cite{carlsson2009topology}, flexibility and rigidity index (FRI) \cite{xia2013multiscale}, and a large number of other network theory models \cite{estrada2020topological}.  Our results are summarized in  Table~\ref{table:descriptors}. The protein folding stability change following mutations is defined by $\Delta\Delta G = \Delta G_{\rm m}-\Delta G_{\rm w}$, where $\Delta G_{\rm w}$ is the free energy change of the wild type and $\Delta G_{\rm m}$ is the free energy change of the mutant type \cite{cang2017analysis}. The folding stability change  $\Delta G$ measures the difference between folded and unfolded states. Thus, a negative $\Delta\Delta G$ means the mutation causing destabilization and vice versa. We use a TDA-based deep learning method \cite{cang2017analysis} to compute $\Delta\Delta G$ in this work. As shown in  Table~\ref{table:descriptors}, a negative folding stability change   $\Delta\Delta G$ is found for the mutation,  indicating a destabilizing mutation L37F for NSP6. Physically, a large hydrophobic residue (F) at the lipid bilayer and solvent interface reduces NSP6 stability. Considering the sequence alignment in Figure \ref{fig:Protter}(c), we have also carried a similar calculation for the mutation of L37V and found its  $\Delta\Delta G =-0.71$ kcal/mol.  Therefore, leucine is favored at residue position 37 for SARS-CoV-2 NSP6. 

Next, the FRI rigidity index $R_\eta$   measures the geometric compactness and topological connectivity of the network consisting of C$_\alpha$ at each residue and the heavy atoms involved in the mutation \cite{xia2013multiscale}. The scale parameter  $\eta$ determines the  range of pairwise interactions.   In this work, the parameter $\eta$ is set to 8\AA. As shown in  Table~\ref{table:descriptors}, the increase in the FRI rigidity index $R_\eta$ is found, indicating 
the increase in the rigidity of the first alpha helix of NSP6. It is known that protein flexibility is required to allow it to function through molecular interactions within the cell, among cells and even between organisms \cite{teilum2009functional}. The quadruplet phenylalanine resulting from mutation L37F can significantly reduce NSP6's interactions with ER. 

\begin{table}[ht]
\centering
\setlength\tabcolsep{1pt}
\captionsetup{margin=0.1cm}
\caption{The folding stability change   and graph network descriptor consisting of wild type and mutant type of NSP6 calculated by web server~\cite{cang2017analysis}. The folding stability change  $\Delta\Delta G = \Delta G_m-\Delta G_w$, where $\Delta G_m$ is free energy change of wild type and $\Delta G_w$ is free energy change of mutation type. $\text{R}_{8}$: FRI rigidity index with $\eta$ of 8; $d$: edge density; $\langle L \rangle$: average path length; $\langle C_b \rangle$: average betweenness centrality; $\langle C_e \rangle$: average eigencentrality; $\langle C_s \rangle$: average subgraph centrality; $\langle M \rangle$: average communicability; $\langle\Theta\rangle$: average communicability angle.}
\begin{tabular}{c|r|r|c}
\hline
\multicolumn{3}{c|}{Mutation L37F}& Destabilizing \\\hline
\multicolumn{3}{c|}{ $\Delta\Delta G$}& -0.74 kcal/mol \\\hline \hline
Descriptors            & Wild type & Mutant type & Relative change (\%) \\\hline 
$\text{R}_8$           & 19.73     & 19.83         &  0.50 \\
$d$                    & 0.05816   & 0.05779       & -0.64 \\
$\langle  L   \rangle$ & 22.7394   & 22.7373       & -0.01 \\
$\langle C_b  \rangle$ & 0.00943   & 0.00942       & -0.11 \\
$\langle C_e  \rangle$ & 0.03508   & 0.03514       &  0.17 \\
$\langle C_s  \rangle$ & 8786858   & 8820927       &  0.39 \\
$\langle  M   \rangle$ & 2826948   & 2822450       & -0.16 \\
$\langle\Theta\rangle$ & 1.8399    & 1.8731        &  1.77 \\ \hline 
\end{tabular}
\label{table:descriptors}
\end{table}

Moreover, seven network models \cite{estrada2020topological} are utilized to analyze the L37F mutation in NSP6 as shown in Table~\ref{table:descriptors}. We consider the network of heavy atoms at residue 37 and  C$_\alpha$ atoms in NSP6. The connections are allowed between any pair of atoms within a cutoff distance of 8.0 \AA.  The mutation-induced decrease in the edge density $d$ is due to the factor that the mutation increases the edge  on the surface of the protein. 
The mutation induced the decrease in the average path length $\langle L \rangle$ is owing to the mutation increases the edge with shorter distances. The mutation induced the decrease in average betweenness centrality $\langle C_b \rangle$ is due to the increase in the crowd of vertices by the mutation residue. The mutation induced the increase in average eigencentrality $\langle C_e \rangle$ is because  increase in the number of the connected components will enlarge the maximal eigenvalues. The mutation induced the increase in the average subgraph centrality $\langle C_s \rangle$ is due to the decrease in the average participating rate of each edge. The mutation induced the decrease in average communicability $\langle M \rangle$ is due to the decrease in the neighbor edges participating rate of each edge. Finally, the mutation induced the increase in average communicability angle $\langle\Theta\rangle$ is due to the increase in the  alignments among different edges.  Together with the FRI rigidity index and the protein folding stability changes, these network assessments show that NSP6 becomes unstable and less functional after the L37F mutation \cite{benvenuto2020evolutionary}. 

\section*{Conclusion}
While the asymptomatic infections of severe acute respiratory syndrome 2 (SARS-CoV-2) have been reported worldwide in the past few months, little is known about the formation mechanism and virological characteristics of asymptomatic infections. This work parses 20656 complete genome sequences of SARS-CoV-2, and for the first time, reveals the relationship between asymptomatic cases and a single nucleotide mutation 11083G$>$T-(L37F) on NSP6. After genotyping 83 sequences with asymptomatic and symptomatic records, we find 11083G$>$T is significantly relevant to the asymptomatic infection. By analyzing the distribution of 11083G$>$T in various countries, we unveil that the 11083G$>$T mutation correlates with the death ratio in Singapore, the United States, the United Kingdom, etc, inferring the correlation between  the asymptomatic infection and the hypotoxicity of SARS-CoV-2. Moreover, we track the dynamics of 11083G$>$T mutation ratio globally  and discover its decaying tendency, indicating that 11083G$>$T  mutation hinders SARS-CoV-2 transmission. Furthermore, the analysis of the distribution of 11083G$>$T in different ages and genders unveils that 11083G$>$T-driven mutation does not have a host-dependent behavior. The protein-specific analysis is also taken into consideration. The 11083G$>$T mutation leads leucine (L) residue at position 37 of NSP6, changing to phenylalanine (F). By employing the graph network analysis and topology-based mutation predictor, we find that 11083G$>$T-(L37F) destabilizes the structure of NSP6 in protein-wise. As one of the most conservative proteins of SARS-CoV-2, NSP6 involves in the viral autophagic regulation. Therefore, this destabilized mutation may result in the inefficiency of NSP6 to participate in the viral protein folding, viral assembly, and replication cycle, which underpins our deduction that 11083G$>$T-(L37F) may be relevant to the asymptomatic infections and weaken the SARS-CoV-2 virulence.  

\section*{Acknowledgment}
This work was supported in part by NIH grant  GM126189, NSF Grants DMS-1721024,  DMS-1761320, and IIS1900473,  Michigan Economic Development Corporation,  George Mason University award PD45722,  Bristol-Myers Squibb, and Pfizer.
The authors thank The IBM TJ Watson Research Center, The COVID-19 High Performance Computing Consortium, NVIDIA, and MSU HPCC for computational assistance.

\section*{Supporting Information}
\subsection*{Additional Analysis}

\subsubsection*{Visualization of  four coronaviral NSP6s in membrane}
\begin{figure}[H]
	\centering
	\includegraphics[width=0.9\textwidth]{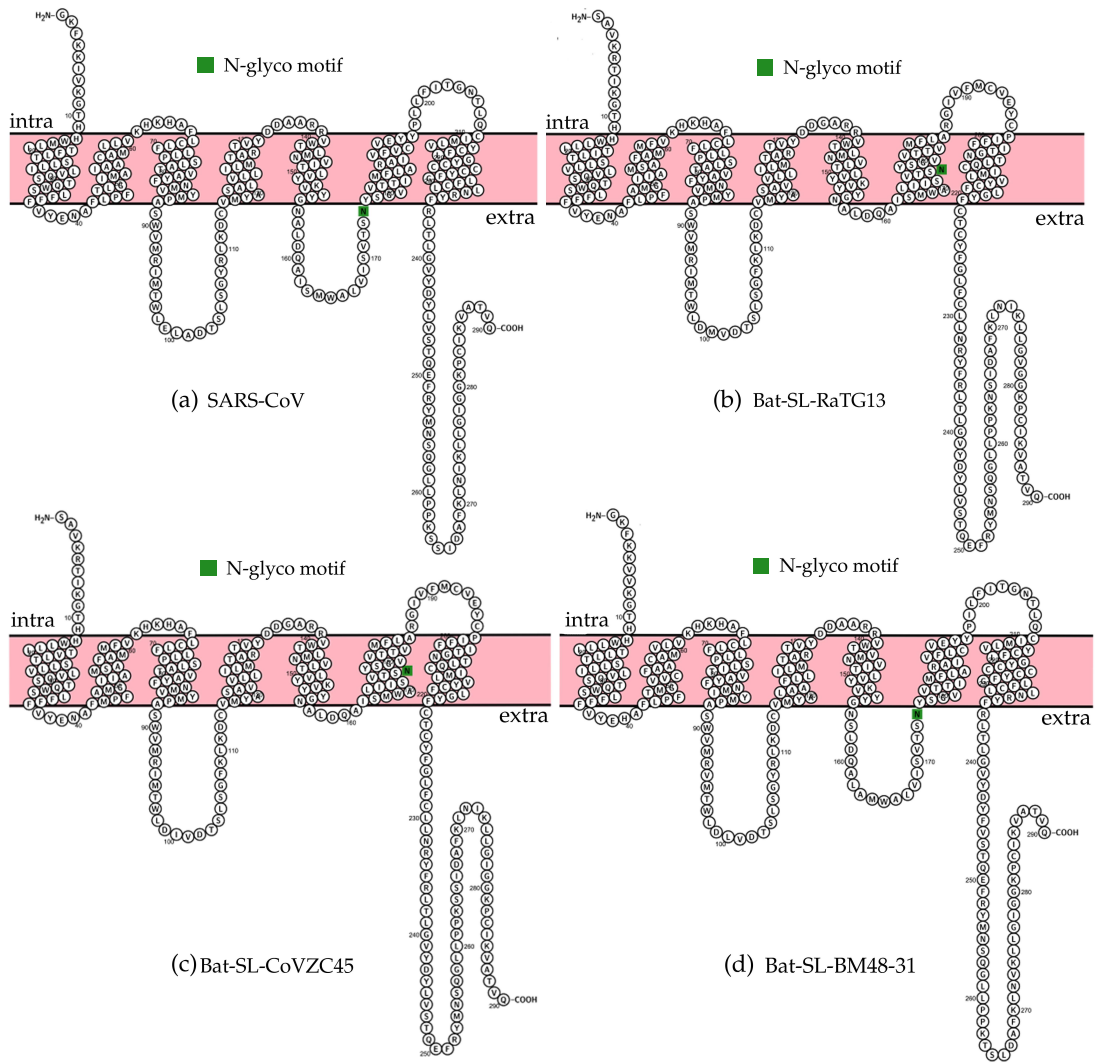}
	\caption{The visualization of  NSP6 proteoforms for a few species close to SARS-CoV-2. (a) SARS-CoV, (b) Bat-SL-RaTG13, (c) Bat-SL-CoVZC45, and  (d) Bat-SL-BM48-31. }
	\label{fig:Protter_Combine}
\end{figure}

\autoref{fig:Protter_Combine} is the visualization of four NSP6 proteoforms of SARS-CoV, Bat-SL-RaTG, Bat-SL-CoVZC45, and Bat-SL-BM48-31. These proteoforms are consistent with that of SARS-CoV-2 described in the main paper, indicating their similar function of regulating cell autophagy. 





\subsubsection*{Network analysis of the SARS-CoV-2 NSP3 L37F mutation }

\begin{figure}[ht!]
	\centering
	\includegraphics[width=0.8\textwidth]{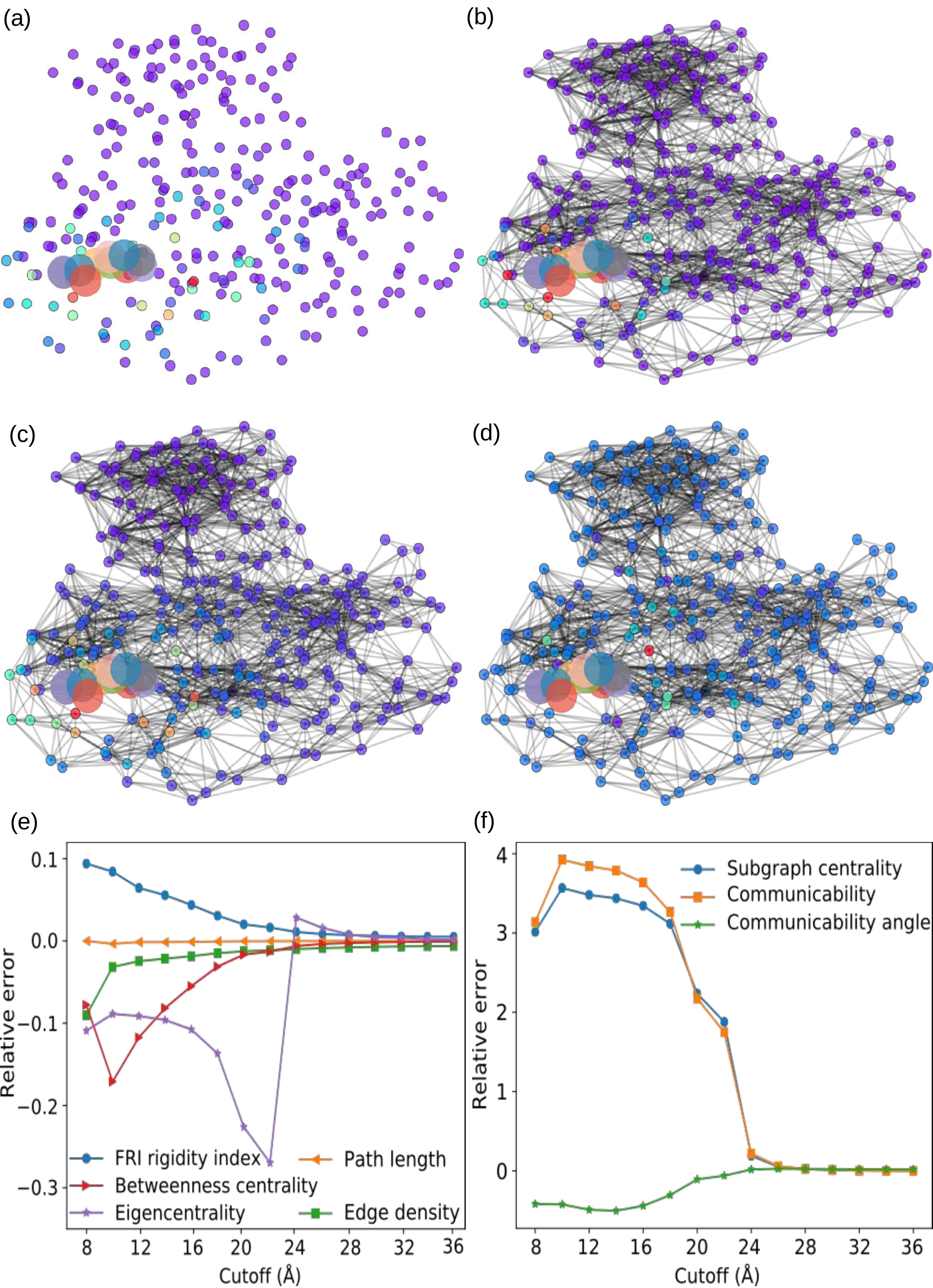}
	\caption{ Network analysis of the SARS-CoV-2 NSP6 L37F mutation. The networks consist of heavy atoms of mutation site 37 and C$_{\alpha}$ atoms of SARS-CoV-2 NSP6. The differences
of descriptors between the network with wild type, leucine, and the network with mutant type, phenylalanie, are displayed. ({a}) FRI rigidity index differences; ({b}) eigencentrality differences; ({c}) subgraph centrality differences; and ({d}) betweenness centrality differences. 
In (e) and (f),  relative changes versus cutoff distances to the mutation site are studied. 
 ({e}) relative changes of  FRI rigidity index, path length, edge density, betweenness centrality, and eigencentrality;  and ({f}) relative changes of subgraph centrality, communicability, and communicability angle.}
	\label{fig:descriptors}
\end{figure}

Figure~\ref{fig:descriptors} shows the network analysis of the SARS-CoV-2 L37F mutation.  
The networks are constructed by heavy atoms at the mutation site and at the C$_{\alpha}$ atoms of SARS-CoV-2 NSP6.  Figures~\ref{fig:descriptors} (a), (b), (c), and (d) present the mutation-induced differences of FRI rigidity index differences,  eigencentrality differences, subgraph centrality differences, and betweenness, respectively. Although these descriptors do not change much globally, their local changes around the mutation site reveal the L37F mutation-induced stress around the mutation site.  
 
Figures~\ref{fig:descriptors} (e) and (f) show the dependence of relative changes over cutoff distances. First of all, every descriptor converges as the cutoff distance increase.    Relative changes do not fluctuate for cutoff distance greater than 24 {\AA }. The relative change of FRI rigidity index monotonically decreases in absolute value as the cutoff increases. Similarly, the edge density has a monotonically decreasing pattern. The two networks are close in the path length descriptor. Interestingly, betweenness centrality descriptor has the largest difference at 10 {\AA } cutoff distance, while the eigencentrality descriptor has the largest difference at 22 {\AA } cutoff distance. For betweenness centrality, it shows that the mutation has the largest impact in the network of C$_\alpha$ within 10 {\AA } to any atoms of the target residue. In Figure~\ref{fig:descriptors} (f), three descriptors show a similar pattern --- they increase in absolute value first and then decrease eventually as the cutoff increases. Overall, the relative change plots indicate that the mutation happening at L37 has the largest impact on the C$_\alpha$ within 10 {\AA } to any atoms of the target residue.

\subsection*{Material and Methods}

\subsubsection*{Data collection and pre-processing}
 
On January 5, 2020, the complete genome sequence of SARS-CoV-2 was first released on GenBank (Access number: NC\_045512.2) by Zhang's group at Fudan University \cite{wu2020new}. Since then, there has been a rapid accumulation of SARS-CoV-2 genome sequences. In this work, 20,656 complete genome sequences with high coverage of SARS-CoV-2 strains from the infected individuals in the world were downloaded from the GISAID database \cite{shu2017gisaid} (\url{https://www.gisaid.org/}) as of June 19, 2020. All the records in GISAID without the exact submission date were not taken into considerations. To rearrange the 20,656 complete genome sequences according to the reference SARS-CoV-2 genome, multiple sequence alignment (MSA) is carried out by using Clustal Omega \cite{sievers2014clustal} with default parameters. 

The amino acid sequence of NSP6 is downloaded from GenBank \cite{benson2009genbank}. The three-dimensional (3D) structure of nonstructure protein 6 (NSP6) in this work is generated by I-TASSER model \cite{yang2015tasser}. The 3D structure graph is created by using PyMOL \cite{delano2002pymol}.

\subsubsection*{Single nucleotide polymorphism genotying}
Single nucleotide polymorphism (SNP) genotyping measures the genetic variations between different members of a species. Establishing the SNP genotyping method to the investigation of the genotype changes during the transmission and evolution of SARS-CoV-2 is of great importance \cite{yin2020genotyping,wang2020decoding}. By analyzing the rearranged genome sequences, SNP profiles, which record all of the SNP positions in teams of the nucleotide changes and their corresponding positions, can be constructed. The SNP profiles of a given SARS-CoV-2  genome isolated from a COVID-19 patient capture all the differences from a complete reference genome sequence and can be considered as the genotype of the individual SARS-CoV-2.

\subsubsection*{Topology-based prediction of protein folding stability changes upon mutation}
In this work, the prediction of NSP6 folding energy changes upon mutation is computed by using the topology based mutation predictor (TML-MP) (\url{https://weilab.math.msu.edu/TML/TML-MP/}) which is briefly reviewed as following and its detail can be found in the literature~\cite{cang2017analysis}. TML-MP applies  element specific persistent homology, which reveals essential biological information \cite{carlsson2009topology,edelsbrunner2000topological}. The method employs the element-specific persistent homology \cite{cang2018integration} and other biological and chemical properties as machine learning features to train gradient boosted regression tree (GBRT) models. The dataset includes 2648 mutations instances in 131 proteins provided by Dehouck et al~\cite{dehouck2009fast}. The error analysis based on the dataset is given as Pearson correlations coefficient ($R_p$) of 0.79 and root mean square error (RMSE) of 0.91 kcal/mol from previous work~\cite{cang2017analysis}.

As the persistent homology widely applied in a variety of practical feature generation problems, it is also successful in the implementation of predictions of protein folding energy changes upon mutation. The key idea in TML-MP is to use the element-specific persistent homology (ESPH) which distinguishes different element types of biomolecules when building persistent homology barcodes. For instance, commonly occurring protein element types include C, N, O, S, and H. However, hydrogen atoms are often absent from PDB data and sulfur atoms are too few in most proteins to be statistically important. Thus, C, N, and O elements are considered on the ESPH in protein characterization. As for persistent homology, barcodes generated based on ESPH provide a topological representation of molecular interactions. Features are extracted from the different dimensions of persistent homology barcodes by dividing barcodes into several equally spaced bins, which is called binned barcode representation. The auxiliary features such as geometry, electrostatics, amino acid types composition, and amino acid sequence are also included for machine learning training. 

The element specific persistent homology is built by adopting the distance function $DI(A_i, A_j)$ describing the distance between two atoms $A_i$ and $A_j$ defined as
\begin{equation}
DI(A_i, A_j) = \begin{cases}
\infty, \text{if Loc($A_i$) = Loc($A_j$),}\\
DE(A_i, A_j), \text{otherwise}, 
\end{cases}
\label{eqn:distance_fcn}
\end{equation}
 
where Loc$(\cdot)$ denotes the location of an atom which is either in a mutant site or in the rest of the protein and $DE(\cdot, \cdot)$ is the Euclidean distance between the two atoms. Then, the persistent homology uses simplicial complexes with a specific rule such as Vietoris-Rips complex, Cech complex, or alpha complex. Vietoris-Rips complex (VC) is used for characterizing first-order interaction where alpha complex (AC) is used for characterizing higher-order patterns. Using ESPH to characterize interactions of different kinds, we construct persistent homology barcodes on the atom sets by selecting one certain type of atoms in mutation site and one other certain type of atoms in the rest of the protein. The set of barcodes from one persistent homology computation as
 $V^{p,d,b}_{\gamma,\alpha,\beta}$ where
\begin{itemize}
	\item $p\in\{\text{VC},\text{AC}\}$ is the complex rule,
	\item $d\in\{DI, DE\}$ is the distance function,
	\item $b\in\{0, 1, 2\}$ is the topological dimensions,
	\item $\gamma\in\{\text{M},\text{W}\}$ is the protein of mutant type or wild type,
	\item $\alpha\in\{\text{C,N,O}\}$ is the element type selected in proteins except in the mutation site,
    \item $\beta\in\{\text{C,N,O}\}$ is the element type selected in the mutation residue.
\end{itemize}
These barcodes are capable of reflecting the molecular mechanism of protein stability. Features are extracted from the groups of persistent homology barcodes. For 18 groups of Betti-0 ESPH barcode such that 9 groups are from the mutant type and 9 groups are from the wild type, one can specify a fixed length interval to divide the ESPH barcodes into a number of equally spaced bins. For example, a length set, $\{[0,0.5],(0.5,1], ..., (5.5,6]\text{\AA}\}$ would turn the 18 groups of Betti-0 ESPH barcode into 18*12 features with dimension of the number of atoms. The death and birth of bars are counted in each bin resulting in features. Therefore, this representation enables us to precisely characterize hydrogen bonds, van der Waals, electrostatic, hydrophilic, and hydrophobic interactions. For the higher-order Betti numbers, the emphasis is given on patterns of both short and long-distance scales. Statistics feature are computed for each group of barcodes for Betti-1 and Betti-2, which are sum, max, and the average of bar length, and max and min of birth and death values. Overall, 12*18 features are generated by Betti-0 on VC, and 7*2*18 features are generated by Betti-1 and Betti-2 on AC.

In TML-MP, gradient boosted regression trees (GBRTs) \cite{friedman2001greedy} are employed to train the dataset according to the size of the training dataset, absence of model overfitting, non-normalization of features, and ability of nonlinear properties. The GBRT method produces a prediction model as an ensemble method which is a class of machine learning algorithms. It builds a popular module for regression and classification problems from weak learners. By the assumption that the individual learners are likely to make different mistakes, the method uses a summation of the weak learners to eliminate the overall error. Furthermore, a decision tree is added to the ensemble depending on the current prediction error on the training dataset. Therefore, this method is relatively robust against hyperparameter tuning and overfitting, especially for data with a moderate number of features. The GBRT is shown for its robustness against overfitting, good performance for moderately small data sizes, and model interpretability. The current work uses the package provided by scikit-learn (v 0.23.0) \cite{pedregosa2011scikit}.

\subsubsection*{Graph network models}
The graph network descriptors are briefly presented which are applied in this work. Graph networks can mimic interactions between pairs of units in molecules. The quantify features of the networks can reveal the biological and chemical properties measured by comparing descriptors on different networks. To detect the single residue impact following mutation, the network consists of a set $S(r)$ of C$_\alpha$ atoms from every residue of protein structure except the target mutation residue where $r$ is the cutoff distance such that a C$_\alpha$ atom is included if it is within $r$ {\AA } to any atom of the target mutation. The total atom set $T(r)$ is defined as the atoms (C, N, and O) of the target residue and C$_\alpha$ atoms of $S(r)$. Moreover, two vertices are connected in the network if their distance is less than 8 {\AA }. Thus the adjacency matrix $A$ can be defined as well where $A$ is a matrix containing 0 and 1 such that $A(i,j)=0$ if $i$-th and $j$-th are disconnected and $A(i,j)=1$ if $i$-th and $j$-th are connected. 

\subsubsection*{FRI rigidity index} FRI rigidity index was introduced to reflect the flexibility between atoms for molecular interaction prediction~\cite{nguyen2016generalized, xia2013multiscale}. The single residue molecular rigidity index measures its influence on the set $S(r)$ which is given as
\begin{equation}
R_\eta = \sum_{i=1}^{N_S}\sum_{j=1}^{N}e^{\big(\frac{\|\textbf{r}_i-\textbf{r}_j\|}{\eta}\big)^2},
\end{equation}
where $N_S$ is the number of C$_\alpha$ atoms of the set $S(r)$ and $N$ is the number of atoms in total atom set $T(r)$. 

\subsubsection*{Edge density}  Edge density is defined based on the adjacency matrix of the total atom set $T(r)$ such as
\begin{equation}
d = \frac{1}{N_S}\sum_{i=1, i\notin I_T}^{N} \sum_{j=1}^{N}A(i,j),
\end{equation}
where $I_T$ is the index set of the mutation residue. 

\subsubsection*{ Average path length} 
Average path length  measures the separation between two vertices of the whole network, which can be used to study infectious diseases spreading in the networks~\cite{watts1998collective}. The average path length for the single mutation system of biomolecular is defined as
\begin{equation}
\langle L \rangle = \frac{1}{2N_S(N-1)}\sum_{i=1, i\notin I_T}^{N} \sum_{j=1}^{N}d(i,j),
\end{equation}
where $d(i,j)$ is the shortest path length between vertices $v_i$ and $v_j$. 

\subsubsection*{ Average betweenness centrality}
Average betweenness centrality  shows communications in a network~\cite{freeman1978centrality}. The average betweenness centrality is given as
\begin{equation}
\langle C_b \rangle = \frac{1}{N_S} \sum_{k=1, k\notin I_T}^{N} \sum_{i=1}^{N}\sum_{j=i+1}^{N} \frac{g^k_{ij}}{g_{ij}},
\end{equation}
where $g^k_{ij}$ is defined as the number of the shortest path between vertices $v_i$ and $v_j$ that passes $v_k$, and $g_{ij}$ is the number of shortest paths between $v_i$ and $v_j$.

 \subsubsection*{  Average egeincentrality} 
Average egeincentrality is the average of elements of the eigenvector $V_{max}$, which is corresponding to the largest eigenvalues of the adjacency matrix~\cite{bonacich1987power} such as
\begin{equation}
\langle C_e \rangle = \frac{1}{N_S} \sum_{i=1, i\notin I_T}^{N} e_i.
\end{equation}

\subsubsection*{Average subgraph centrality}
Average subgraph centrality is built on the exponential of the adjacency matrix, $E=e^A$. The subgraph centrality is the summation of weighted closed walks of all lengths starting and ending at the same node~\cite{estrada2005subgraph, estrada2020topological}. Thus the average subgraph centrality reveal the average of participating rate of each vertex in all subgraph, which is given as
\begin{equation}
\langle C_s \rangle = \frac{1}{N_S} \sum_{i=1, i\notin I_T}^{N} E(i,i).
\end{equation}

\subsubsection*{ Average communicability}
Average communicability  is defined in a similar way as the subgraph centrality on the exponential of the adjacency matrix~\cite{estrada2008communicability, estrada2016communicability,estrada2020topological}, which is
\begin{equation}
\langle M \rangle = \frac{1}{N_S(N-1)}\sum_{i=1, i\notin I_T}^{N}\sum_{j=1}^{N} E(i,j), 
\end{equation}

\subsubsection*{Average communicability angle}
Average communicability angle is given by  \cite{estrada2016communicability}
\begin{equation}
\langle \Theta \rangle = \frac{1}{N_S(N-1)}\sum_{i=1, i\notin I_T}^{N}\sum_{j=1}^{N} \theta(i,j),
\end{equation}
where $\theta(i,j) = \cos^{-1} \Big( \frac{E(i,j)}{\sqrt{E(i,i),E(j,j)}} \Big)$.

\bibliographystyle{abbrv}
\bibliography{refs}

\end{document}